\documentstyle[aps,epsf]{revtex}
\def\l{{\bf l}}
\def\q{{\bf q}}
\def\k{{\bf k}}

\def\pbar{\rlap\slash p}
\def\lbar{\rlap\slash l}
\def\kbar{\rlap\slash k}

\begin{document}

\begin{flushright}
JLAB-THY-97-32 \\
\end{flushright}

\vspace{2cm}

\begin{center}
{\Large \bf Consistent Analysis of  $O(\alpha_s)$ Corrections to Pion 
 Elastic Form Factor}
\end{center}
\begin{center}
{ADAM SZCZEPANIAK\footnotemark, CHUENG-RYONG JI 
} \\ 
 {\it  Department of Physics,
   North Carolina State University,
   Raleigh, NC 27695-8202 }\\
   ANATOLY RADYUSHKIN\footnotemark  \\
 {\it  Physics Department, Old Dominion University, Norfolk, VA
23529  }\\   {\it and} \\
  {\it  Jefferson Lab, Newport News, VA
23606 }
\end{center}
\vspace{2cm}

\footnotetext[1]{Present address: Department of Physics, 
Indiana University, Bloomington, IN 47405}

\footnotetext{Also Laboratory of Theoretical Physics, 
JINR, Dubna, Russian Federation}

\begin{abstract}

 We examine the role of  $O(\alpha_s)$ perturbative corrections 
to the pion elastic form factor $F_\pi(Q^2)$.  
 We express the quark current three-point function 
 in terms of light cone variables and use Borel 
transformation to simultaneously model 
the Feynman mechanism 
 contribution  determined by 
the soft part of the pion light cone wave function  and the hard term    
involving  one gluon exchange. We find that for 
 $Q^2 \sim 4 \mbox{ GeV}^2$ the total $O(\alpha_s)$   
 contribution  may reach 
$30\%  $ of the soft contribution, 
even though its hard,  factorization scale dependent 
 part  
 remains rather  small.  
\end{abstract}

\vspace{1cm}


\newpage

\section{Introduction}

The interplay between contributions due to
Feynman \cite{feynman} and hard-scattering 
scenarios \cite{hard},  the two basic 
mechanisms  determining the large-$Q^2$ behavior of hadronic form
factors, is a crucial  problem \cite{ils,rad} 
for studies of exclusive 
processes in quantum chromodynamics.  
To make  a  meaningful   comparison of their magnitude, 
one should be able to calculate both contributions
within the same self-consistent framework. 
 To implement such a program,  and especially in order to be able 
to  take into account effects due to the nonperturbative
part of the  hadronic wave functions  it is 
natural to rely on the   
  light cone  quantization \cite{Lep80}  which  enables a 
   Fock space representation for the hadronic current matrix elements. 
 An important issue which has to be addressed 
  when studying exclusive amplitudes in the 
 region where both hard and soft QCD processes 
 are important, is how to implement  
 factorization,  $i.e.,$  separate perturbative contributions from those 
  intrinsic to the bound-state wave function \cite{Lep80,ils,rad}.
 In the light cone quantization, 
the pion form factor can (in principle)
 be determined from
\begin{equation}
F_{\pi} (Q^2) = \hat\Psi \otimes \hat\Psi  
=\sum_n\int [dx_i d\k_{i\perp}]_n  
\hat\Psi_n(x_i,\k_{i\perp})
\hat\Psi_n(x_i,\k_{i\perp}+\delta_i\q_\perp), \label{fpi0}
\end{equation} 
where the summation extends over all quark/gluon Fock sectors 
which have  a   
nonvanishing overlap with the pion, $\hat\Psi_n$ 
are the corresponding wave functions, 
 $[dx_i d\k_{i\perp}]_n$ is the relativistic measure within the 
 $n$-particle sector and 
 $\delta_i=(1-x_i)$ or $-x_i$ depending on whether $i$ refers to 
 the  struck 
quark or a spectator, respectively. 
 Instead of having to deal with an infinite set of wave 
 functions,  $\hat\Psi_n$ which describe 
both the low and high momentum partons 
one can  use  the factorization procedure which allows   
to express high momentum 
 tails of these wave functions  in terms of suitably defined soft 
 wave functions  
 $\Psi_n$ with nonvanishing support only at low momenta. 
 In principle,  this may be achieved by constructing an  
 effective Hamiltonian which has nonvanishing matrix elements   
  within the low momentum subspace.  This effective Hamiltonian 
 incorporates  couplings between low and high momentum degrees of 
 freedom  present in the bare Hamiltonian, through a set of 
 effective potentials connecting the low momentum states. 
 The eigenstates of the effective 
Hamiltonian determine soft hadronic wave functions, $\Psi_n$.  
 Because of asymptotic freedom, the removal of couplings to high momentum 
 Fock states and construction of effective operators can be performed  
 in a power series expansion in $\alpha_s$.  
  A similar procedure when applied to the current $\bar q \gamma^+ q$ yields 
 an effective 
current  
operator.   
 The contribution from the lowest
  order one-body current, 
 as given by Eq.~(\ref{fpi0}) is modified
in 
higher orders by contributions 
 from two-body and 
 more complicated operator matrix elements resulting in 
\begin{eqnarray}
& & F_{\pi} (Q^2) = \Psi \otimes T \otimes \Psi  \label{fpieff} \\ 
& & \equiv \sum_{nm}\int [dx_i d\k_{i\perp}]_{n} [dy_i d\l_{i\perp}]_{m}
\Psi_{n}(x_i,\k_{i\perp};\mu)
T_{nm}(x_i,\k_{i\perp}; y_i,\l_{i\perp};\q_{\perp};\mu) 
\Psi_{m}(y_i, \l_{i\perp};\mu),
\nonumber  
\end{eqnarray}
where summation over $n,m$ now extends over the low momentum states only, 
  and 
 $T_{nm}$ are the partonic matrix elements of the effective current 
 operator. 
 The dependence on the 
scale 
separating low and high momenta  
is indicated  by   $\mu$. The matrix elements, $T_{mn}$   
 mix low momentum partons described by the wave functions, $\Psi$ 
  with intermediate high 
 momentum states. The free energy of these intermediate states is 
 bounded from below by $\mu$ and in order for the perturbative 
  approach to make sense  
 $\mu$ has to be chosen to be much larger than $\Lambda_{QCD}$ so that
  $\alpha_s=\alpha(\mu)$  is small. 
  At large momentum transfer we may
ignore the dependence of $T_{mn}$   on  small  transverse momenta  
 $\k_\perp$ and $\l_\perp$ present in the soft wave functions.  
After that,  $\Psi_n (x_i,\k_{i\perp}; \mu )$   
and $\Psi_{m}(y_i, \l_{i\perp};\mu)$  
can be integrated over $\k_\perp$ and $\l_{\perp}$, respectively  
to produce the distribution amplitudes ($e.g.,$  $\phi(x,\mu^2)$   
for the pion)  depending only on   longitudinal momenta.    
To avoid large logarithms $O(\alpha_s \log(Q^2/\mu^2))$ emerging    
 from the effective  current matrix elements, we can set   
$\mu \sim |\q_\perp| = \sqrt{Q^2}$, absorbing the logarithms    
 into the evolution of the  distribution  amplitudes, $\phi(x,Q^2)$.   
In particular,  the latter  enter the well-known   
formula for the pion form factor \cite{tmf,Lep80,CZ}:   

\begin{equation}
F(Q^2) = {{16\pi C_F \alpha_s(Q^2) }\over Q^2} 
\left[ \int  {{\phi(x,Q^2)}\over 
 x} \, dx  \right]^2 \, . \label{fpiass}
\end{equation}  
 In the light cone gauge $A^+=0$,  the
distribution  
amplitudes are determined by 
 matrix elements 
of composite operators in which the covariant derivatives   
$D^+$ coincide with the ordinary  $\partial^+$ ones.  
 In other words,  to 
the  
 leading order in $\alpha_s$ and $1/Q^2$ ,
  only the minimal,  valence Fock sector contributes. 
 Conversely,  any attempt to improve 
the leading order formula, 
 for example by   
 keeping the transverse momentum dependence of $T_{nm}$, requires 
to take into account the    
 presence of nonvalence degrees of freedom. 
 
Another important observation, 
supported by QCD sum rules\cite{IoS,NeRa82} and  
by  various  
constituent quark   
model studies
\cite{cqm,sch,cqm1}, is that   
the soft
contribution to the form factor may be 
  quite substantial.  
To make a self-consistent  comparison of  the relative size  
of the purely soft  term and those  
containing the hard gluon exchange,  
one should be able to calculate  them using the same 
formalism,  which is  not necessarily based on the 
 $1/Q^2$ expansion  but which rather enables to compute the perturbative 
 corrections to the soft form factor.  In other words, 
we propose  
to take advantage 
 of the formula given in Eq.~(\ref{fpieff}),
 retaining there the  
soft wave  functions, rather than 
expressing 
the form factor in 
terms of  
the distribution amplitudes  
as given in Eq.~(\ref{fpiass}) for the leading 
contribution. 

In practice, to develop a 3-dimensional renormalization 
 program for a Hamiltonian which also produces light-cone bound states  
 is rather difficult. One of the reasons 
is  
that cut-offs violate 
 Lorentz and gauge symmetry, and thus 
it is  highly nontrivial  
to achieve the necessary 
 cancellation of IR divergences. 

  It is also unclear how  one could systematically 
incorporate  
the pQCD  
radiative  corrections 
into a bound-state  
approach  with phenomenological 
 low-energy 
interaction  
kernels.   

A simple approach which enables the connection between  
 the three-dimensional soft wave functions and the high energy scattering kernels  
 of perturbative QCD  was  proposed in Ref.~\cite{b2pi} where it was 
 also  applied to describe the  
  soft and hard contributions to the heavy-to-light meson transition 
 form factor. 
 The method relies on Green's 
 functions 
 just like    
the QCD sum rule approach \cite{SVZ}. 
Let us recall that   
the   
QCD sum rule method  
uses 
the  
  quark-hadron duality 
relation  

\begin{equation}
\int_0^{\infty} \rho^{\rm hadron}(s)e^{-s/M^2} \, ds 
= \int_0^{\infty} \rho^{\rm quark}(s)e^{-s/M^2} \, ds  + 
\sum_{N} \frac{\langle {\cal O}_N \rangle}{(M^2)^N}, \label{QCDSR}
\end{equation}  
between the hadronic spectral density $\rho^{\rm hadron}(s)$
on one side and the perturbative  spectral density 
$\rho^{\rm quark}(s)$ and the condensates $\langle {\cal O_N} \rangle$
on the other. The exponential weight $e^{-s/M^2}$ 
in this relation results from the Borel transformation
which both emphasizes the lowest resonance contribution
into the dispersion integral and 
improves the convergence properties  of the condensate series
generated by the operator product expansion.
For an appropriate  choice  of the Borel parameter
$M$, the left-hand side  of the duality relation
is rather closely approximated by   the lowest 
resonance contribution while the right-hand side 
is essentially given by the perturbative term alone.
In such a situation,  one can approximate the lowest resonance 
contribution by the Borel transform of the 
relevant  Green's function. 
Such an approximation may be  rather accurate 
when the Borel parameter is taken within 
 a specific range  for which the higher state 
contribution into   the  left-hand side  is  close in 
magnitude to the condensate contribution to the  right-hand side. 
As we will see below, such an ansatz  is analogous to
the exponential (oscillator-type) 
 model for the soft pion wave function.
Calculating  the $\alpha_s(M^2)$ expansion for the 
Green function ($i.e.,$ higher-order 
terms in  $\rho^{\rm quark}(s)$),
we can  construct 
a unified self-consistent model 
which includes both 
 the lowest
Fock state contribution   and the higher
Fock terms generated by  the 
radiative corrections.   
The model, e.g., automatically preserves
such an important property as gauge invariance.
The latter is crucial  for cancellation
of infrared divergences.

In application to form factors, as described  
in Ref.~\cite{b2pi}, the  
  Borel transformation is used as a method
 which,  on a loop-by-loop basis, enables  to 
explicitly  
identify  the wave-function-like contributions 
to the form factor which have the structure 
of the Fock-space decomposition  
 Eq.~(\ref{fpieff}). 
    An important advantage
 of this method over a naive summation of perturbative 
 corrections  to model wave functions, 
is that no {\it ad hoc} assumptions about 
how to connect soft and hard regimes have to be made. The formalism 
constrains how 
the  
IR divergences are distributed among 
the wave functions and scattering kernels to       
 guarantee the IR finiteness and factorization scale independence 
of a matrix element.  
As we will see,  this requires 
that both the  higher Fock space contributions, 
 $\Psi_{n}$, with,  e.g.,  $n=\bar q q g$,  
 and the radiative corrections to the current matrix elements, 
 $T_{nm}$ 
should be  
present.  

The paper is organized as follows. In the next section we discuss the model and details of 
the  form factor calculation. The Sudakov suppression of the electromagnetic vertex 
 is discussed in Sec.III. Conclusions are summarized in Sec.IV.

\section{Model Specification}

As an illustration of the method,
consider a two-point amplitude defined by 

\begin{equation}
(p^+)^2 \Pi (p^2) = -i \int d^4x e^{ipx} 
\langle 0|T\{ \bar{q}(x) \gamma^+ \gamma_5 q(x), \bar{q}(0) \gamma^+ \gamma_5 
 q(0)\}|0\rangle.
\end{equation}
After the Borel transformation,

\begin{equation}
\Pi(p^2) \rightarrow \Pi(\beta) = \frac{1}{2 \pi i} 
\oint dp^2 e^{-p^2/2\beta^2} \Pi (p^2) = \int ds e^{-s/2\beta^2}
 \rho(s),
\end{equation}
where $\rho(s) \equiv (\Pi(s-i\epsilon)-\Pi(s+i\epsilon))/2\pi i$, 
 the bare quark loop contribution to $\Pi$ yields, 

\begin{equation}
\Pi^{\rm quark}(\beta) =  \int ds e^{-s/2\beta^2}
 \rho^{\rm quark}(s) = 
4\times 2\times 3\int [dx_i d\k_{i \perp}]_{2} \
\exp\biggl[-{1\over {2\beta^2}} \sum_{i=1}^{2}\frac{\k^{2}_{i\perp}+
m_{i}^2}{x_{i}}\biggr]\ ,
\end{equation}
 with the numerical factors in front of the integral 
 coming form the trace over Dirac indices, the quark 
current matrix element  $\bar u(\lambda)\gamma^+\gamma_5 v(\lambda') = 
2\delta_{\lambda,-\lambda'}$, 
 and sum over three colors respectively; the measure is given by 
\begin{equation}
 [dx_i d\k_{i \perp}]_{2} = 16\pi^3\delta \left (1-\sum_i x_i \right )
\delta \left (\sum_i \k_{i\perp}  \right ) \prod_i {{dx_i d^2\k_{i\perp}}\over {16\pi^3}}.
\end{equation}
The quark spectral function, $\rho^{\rm quark}$ is given by 
\begin{equation}
\rho^{\rm quark}(s) = {2\sqrt{6}}\int [dx d^2\k_\perp]_2  \, {2\sqrt{6}} \, 
\delta\left(s - \sum_i^2{{\k_{i\perp}^2+m_i^2}\over {x_i}}\right). \label{rhoq}
\end{equation}
The contribution of the lowest hadronic  state  to  
$\Pi(\beta)$ is given by  
\begin{equation}
\Pi^{\rm hadron}(\beta) = f_h^2 \exp\biggl[-{M_h^2\over {2\beta^2}}\biggr],
\end{equation}
where in our case $M_h = M_\pi$ is the mass of the ($J^{P}=0^-$) ground state, the  pion,  
and $f_h = f_\pi$ is the pion decay 
constant 
\begin{equation}
\langle 0| \bar q(0)\gamma^+\gamma_5 q(0|P \rangle = if_\pi P^+ \, .
\end{equation}
Therefore, if we assign
\begin{equation}
\Psi_2(x_{i},k_{i\perp}) = {{2 \sqrt{6}}\over {f_\pi}} 
\exp\biggl(\frac{1}{2\beta^{2}}\biggl[M_\pi^2 - \sum_{i=1}^{2}\frac{{\bf 
k}^{2}_{i\perp}+ m_{i}^2}{x_{i}}\biggr]\biggr) , \label{wfpi1}
\end{equation}
then the decay constant $f_\pi$ may be expressed 
in terms of this valence wave function: 
\begin{equation}
f_\pi = 2 \sqrt{6} \int [dx_i d\k_{i \perp}]_{2} 
\Psi_2 (x_i, k_{i \perp}).  \label{fpi}
\end{equation}
This expression agrees with the one obtained using  the
 light cone quantization directly. 
In the following we will drop the subscript on the wave function and 
 on the light cone measure 
  when referring to the valence  
 $\bar q q$ component of the ground state wave function. In the valence sector,  
the individual quark momenta $x_i,\k_{i\perp}$ may be replaced by the Jacobi 
variables defined by $x = x_1$, $x_2 = 1 - x_1$ and 
$\k_\perp = \k_{1\perp} = - 
\k_{2\perp}$ such that 
\begin{equation}
[dx_i d\k_{i\perp}]_2 = [dx d\k_\perp] = {{dx d^2\k_{\perp}}\over {16\pi^3}}.
\end{equation}

In addition to   the Borel transformation we will  use  also 
 another  smearing procedure corresponding to 
the ``local duality'' prescription. 
Its origin can be briefly explained in the following way.
 In QCD sum rule calculations one typically uses the following 
ansatz for the hadronic spectral 
function, $\rho^{\rm hadron}(s)$  

\begin{equation}
\rho^{\rm hadron}(s) \approx 
 f_h^2 \delta(s - M_h^2) + \rho^{\rm quark}(s)\theta(s - s_0),
\end{equation}
with the effective continuum threshold $s_0$ fitted from the 
relevant QCD sum rule Eq.~(\ref{QCDSR}) by requiring the most stable 
[$M^2$-independent] result for physical quantities.
In many cases, the stability persists in the whole  large-$M^2$ 
region.  Fitting $s_0$ and then taking the limit 
 $M\to \infty$ for the Borel parameter in  Eq.~(\ref{QCDSR})
gives~\cite{ARCracow}

\begin{equation}
f_h^2 = \int_0^{s_0} \rho^{\rm quark}(s)\,  ds.
\end{equation}
The local duality states that the two densities, $\rho^{\rm hadron}$ and $\rho^{\rm quark}$ 
 give the same result provided the duality interval $s_0$ is properly chosen.  
 Comparing Eqs.~(\ref{rhoq}) and ~(\ref{fpi}) results in 

\begin{equation}
f_\pi = 2\sqrt{6}\int [dx_i d^2\k_{i\perp}]_2 \Psi^{\rm LD}(x_i,\k_{i\perp}),
\end{equation}
where the ``local duality'' pion wave function $\Psi^{\rm LD}$ is defined by 

\begin{equation}
\Psi^{\rm LD}(x_i,\k_{i\perp}) = {{2\sqrt{6}}\over {f_\pi}}\theta\left(s_0 - 
 \sum_{i}^2{{\k_{i\perp}^2 + m_i^2}\over {x_i}}\right). \label{pld}
\end{equation}
The parameter  $s_0 = 4\pi^2 f^2_\pi 
\sim (830\mbox{MeV})^2$ serves here as a transverse momentum cutoff and plays a role similar 
to the wave function scale $\beta$ 
in the approach based directly on the Borel transformation. 
 In the following,   we will continue using the Borel
 transformation when deriving analytical expressions.
  But we  will also give numerical results obtained for the local duality 
 wave function.

\subsection{Soft contribution}
To obtain a 
model expression for the  
soft contribution to $F_\pi(Q^2)$ in a  form of an integral 
over the valence light cone wave function,
 consider the three-point function
 
 \begin{equation}
(p^+)^3 T(p^2,p'^2) = -\int d^4x d^4y e^{ipx-ip'y}
  \langle 0| T\bar q(x) \gamma^+ \gamma_5 q(x),
 \bar q(0) \gamma^+ q(0), \bar q(y) \gamma^+ \gamma_5 q(y) 
| 0 \rangle   \label{tpf}
\end{equation}
in the kinematical region defined by $p'^+ - p^+ = q^+ = 0$ and
 $(p'-p)^2 = q^2 = - \q_\perp^2 =  -Q^2$. 
The quark loop contribution to the double Borel transform ation,

\begin{equation}
T(p^2,p'^2)\to T(\beta) =
\int dsds'  e^{-s/2\beta^2} e^{-s'/2\beta^2}
\rho(s,'s),\label{bt2}
\end{equation}
is 
\begin{equation}
\rho^{\rm quark}(s,s') = Sp T(s,s')
= 2(2\sqrt{6})^2\delta\left(
s - \sum_i^2 {{\k_{i\perp}^2 + m_i^2}\over {x_i}}\right)
\delta
\left(s' - \sum_i^2 {{(k_{i\perp} +\delta_i\q_\perp)^2 + m_i^2}
\over {x_i}}\right),
\end{equation}
and therefore the $q\bar q$ contribution yields
\begin{equation}
T^{\rm quark}(\beta)
= 2 f_\pi^2 \exp \left(- {{2 M_\pi^2}\over {2\beta^2} } \right)  
\int [dx d\k_\perp]
 \Psi(x,\k_\perp)\Psi(x,\k_\perp - x \q_\perp). \label{t1}
\end{equation}
The ground state hadronic contribution to $T$  
is given by 
\begin{equation}
T^{\rm hadron}(\beta) = 2 f_\pi^2F_\pi(Q^2) \exp \left
( - {{2 M_\pi^2}\over {2\beta^2} } \right).
\label{t2}
\end{equation}
Identifying  
Eq.~(\ref{t2}) and Eq.~(\ref{t1})
yields the standard expression for the 
 valence quark wave function contribution to $F_\pi (Q^2)$,
\begin{equation}
F_\pi(Q^2) =   \int [dx d\k_\perp]
 \Psi(x,\k_\perp)\Psi(x,\k_\perp - x \q_\perp). \label{soft}
\end{equation}

 As illustrated on the two examples above, for a single quark loop 
 we have  generated expressions for hadronic matrix elements
  in terms of 
model   
light cone wave functions.
 As will be shown below, this remains to be the case
  beyond the single quark loop.
When the higher loops  are considered, the   
 nonvalence wave functions   emerge.

 In the following we shall work in the chiral limit and set 
$M_{\pi}=0$. With the local duality wave function given by  Eq.~(\ref{pld}),  
 the normalization condition for the form factor, $F_\pi(0)=1$ is satisfied 
automatically.  Taking the  wave function (\ref{wfpi1}) based on the Borel transformation, 
 produces only  a half of the necessary form factor normalization
at $Q^2=0$.  In the standard light-cone approach, 
the remainder comes from contributions due to the higher Fock
components like  $\bar qG \ldots G q$ corresponding to the ``intrinsic'' 
gluon   content of the hadron.  
The intrinsic gluons should  be contrasted with 
 those  emerging from explicit
radiative  corrections to the
two-body wave function. 
Since the local duality wave function
provides 100\%   of the $Q^2=0$ normalization,
one can interpret it as an {\it effective}
two-body wave function absorbing in itself  the 
soft part of the higher-Fock states.
In the spirit of effective wave functions,
we will also modify the Gaussian wave function and 
instead of Eq.~(\ref{wfpi1}) we will use

\begin{equation}
\Psi_2(x_{i},k_{i\perp}) = {{2 \sqrt{6}}\over {f_\pi}} N
\exp\biggl(\frac{1}{2\beta^{2}}\biggl[M_\pi^2 - \sum_{i=1}^{2}\frac{{\bf 
k}^{2}_{i\perp}+ m_{i}^2}{x_{i}}\biggr]\biggr) , \label{wfpi}
\end{equation}
with  the extra normalization constant $N$ and the parameter 
$\beta$  fixed by 
Eq.~(\ref{fpi}) and $F_\pi(0)=1$ which lead 
 to $N=2$, $\beta = \pi f_\pi\sim 400\, \mbox{MeV}$.

\subsection{One-gluon exchange and  hard contribution}

The one-gluon-exchange diagrams 
are shown in Fig.~1b-c. In addition, to order 
$\alpha_s$ there are three self-energy type diagrams from dressing of the 
propagators  of the triangle diagram shown in Fig.~1a.
 The hard-gluon-exchange  contribution to $F_\pi$ will be defined as an IR-regular 
part of two  diagrams (Figs.~1c,d) corresponding to
  vertex corrections  to  the axial vector 
currents. 
 The IR finite  part of the diagram with electromagnetic  
vertex correction  (Fig.~1b)  contains terms which 
produce  the Sudakov  
suppression after all-order summation.  
 Since the operators which define the correlator 
 $T$ correspond to 
conserved currents, 
the sum 
 of all  six amplitudes is UV finite. It is also IR finite 
 since there is no net 
 color flowing into  or out of the triangle. 

 Just like in   case of the  one-loop triangle diagram we calculate 
  Feynman amplitudes corresponding to each  two-loop diagram 
  by performing analytically the integrals over the 
 ``minus'' light cone components of loop momenta. 
Then, we apply the Borel transformation, 
 and the net result may be schematically written as 
\begin{equation} 
F_\pi = \Psi \otimes  I  \otimes  \Psi +\Psi \otimes T  \otimes  \Psi
\Psi_{g} \otimes T_{g1} \otimes  \Psi
+ \Psi \otimes T_{g2} \otimes \Psi_{g}
+ \Psi_{g} \otimes I_{gg} \otimes \Psi_{g}.
\label{vertex}
\end{equation}
The first term comes from the one-loop diagram. The remaining four terms 
represent the $O(\alpha_s)$ contributions.
The $T$-term corresponds to the vertex correction to the 
electromagnetic current convoluted with two $\bar qq$, $i.e.,$  
 valence meson wave
functions.   Two other amplitudes,  $T_{g1}$ and  
$T_{g2}$,  are convoluted with the 
 valence  wave function on one end and the 
 nonvalence  $\bar q q g$ wave function on the other one.
Finally, the $I_{gg}$-term corresponds to 
 the  nonvalence 
 $\bar q q g$ meson wave functions induced both in the initial
and final states.

 As an example consider the diagram shown in Fig.~1b. 
 The Feynman amplitude is given by
\begin{eqnarray}
(p^+)^3 T(p^2,p'^2) & = & 12\pi C_F\alpha_s 
\int {{d^4k}\over {(2\pi)^4}} {{d^4l}\over {(2\pi)^4}} \nonumber \\
& & \times \mbox{Tr} \left[ {
{ \gamma^+\gamma_5(\pbar'-\kbar)\gamma^\alpha(\pbar'-\lbar)\gamma^+
(\pbar-\lbar)\gamma^\beta(\pbar-\kbar)\gamma^+\gamma_5(-\kbar) }
\over
{(p'-k)^2(p'-l)^2(p-l)^2(p-k)^2k^2(k-l)^2}
} \right].\label{qcd}
\end{eqnarray}
The $k^-$ and $l^-$ integrals pick up poles in the spectator quark ($1/k^2$) 
and gluon ($1/(k-l)^2$) propagators leading to

\begin{eqnarray}
T(p^2,p'^2) & = & 2(8\pi C_F\alpha_s) (2\sqrt{6})^2
\int {{dx d\k_\perp}\over {16\pi^3}}
     {{dy d\l_\perp}\over {16\pi^3}}
{
 { y[\l_\perp - \q_\perp]\cdot[\l_\perp + (1-y)\q_\perp] }
 \over {\l_\perp^2[\l_\perp-y\q_\perp]^2} }
 \nonumber \\
& & \times \left[
  {1 \over {p^2 - {{\k_\perp^2}\over {x(1-x)}} + i\epsilon}}
 - {1 \over {p^2 - {{\k_\perp^2}\over {x(1-x)}}
                  - {{\l_\perp^2}\over {y(1-y)(1-x)}} + i\epsilon}}
           \right] \nonumber \\
& & \times
           \left[
  {1 \over {p'^2  - {{(\k_\perp-x\q_\perp)^2}\over {x(1-x)}} + i\epsilon}}
 - {1 \over {p'^2  - {{(\k_\perp-x\q_\perp)^2}\over {x(1-x)}}
                  - {{(\l_\perp-y\q_\perp)^2}\over {y(1-y)(1-x)}} 
+ i\epsilon }}
            \right].
\label{qcd2}
\end{eqnarray}
After the Borel transformation,  the contribution to $F_\pi$ from this diagram 
 can be expressed in a form of Eq.~(\ref{vertex}). Explicitly, 
\begin{eqnarray}
 & & F_\pi(Q^2) \to F_\pi^\Gamma(Q^2) = 
 \int {{dx d\k_\perp}\over {16\pi^3}} \Psi(x,\k_\perp)T^\Gamma
\Psi(x,\k_\perp-x\q_\perp)
 \nonumber \\
 & & + \int {{dx d\k_\perp}\over {16\pi^3}}{{dy d\l_\perp}\over {16\pi^3}}
\left[\Psi_g(x,\k_\perp;y,\l_\perp)T_{g1}^\Gamma \Psi(x,\k_\perp-x\q_\perp)
 + \Psi(x,\k_\perp)T_{g2}^\Gamma
\Psi_g(x,\k_\perp-x\q_\perp;y,\l_\perp-y\q_\perp) \right.\nonumber \\
& &  \phantom{  F_\pi^\Gamma(Q^2) = \int {{dx d\k_\perp}\over {16\pi^3}}{{dy
d\l_\perp}\over {16\pi^3}}
} \left.
  +\Psi_g(x,\k_\perp;y,\l_\perp)
I_{gg}^\Gamma\Psi_g(x,\k_\perp-x\q_\perp;y,\l_\perp-y\q_\perp)\right],
\label{vert}
\end{eqnarray}
where
\begin{equation}
 T^\Gamma =
 8\pi C_F\alpha_s  \int {{dy d\l_\perp}\over {16\pi^3}}
{1\over {y(1-y)}}
{{N(y,\l_\perp)}\over {D(y,\l_\perp)D(y,\l_\perp - y\q_\perp)}},
\end{equation}
\begin{equation}
 T^\Gamma_{g1} = - \,{{\sqrt{8\pi C_F \alpha_s}} \over {y(1-y)}}
{{N(y,\l_\perp)}\over {D(y,\l_\perp)}},\;\;
 T^\Gamma_{g2} = - \, {{\sqrt{8\pi C_F \alpha_s}} \over {y(1-y)}}
{{N(y,\l_\perp)}\over {D(y,\l_\perp - y\q_\perp)}}, \;\;
I_{gg} = {{N(y,\l_\perp)}\over {y(1-y)}},
\label{tsg}
\end{equation}
 and we have defined 
\begin{equation}
N(y,\l_\perp) \equiv {{y[\l_\perp - \q_\perp]\cdot [\l_\perp + (1-y)\q_\perp]}
\over {y(1-y)}},\;\; D(y,\l_\perp) \equiv {{\l_\perp^2}\over {y(1-y)}}.
\label{nd}
\end{equation}
The valence wave function  $\Psi$ is 
 given by  Eq.~(\ref{wfpi}) (with $M_{\pi}=0$). The contribution from the 
 $\bar q qg$ state in the electromagnetic vertex loop is determined by 
 $\Psi_g$ which is given by
\begin{equation}
\Psi_g(x,\k_\perp;y,\l_\perp) 
= {{\sqrt{8\pi\alpha_s C_F}}\over {D(y,\l_\perp)}}
{{2\sqrt{6}}\over f_\pi}
\exp\left[{1\over {2\beta^2}}\left( - {\k_\perp^2\over 
{x(1-x)}}
 - {\l_\perp^2 \over {y(1-y)(1-x)}} \right) \right].
\end{equation}
Here $y$ represents the fraction of longitudinal momentum of a 
 quark carried away by the gluon. The spectator quark has longitudinal 
 momentum $x$.    
The transverse momentum  $\l_\perp$ is 
  the relative momentum between the struck quark and the gluon.  
  Note,  that  just like the argument of the valence 
 wave function  $\Psi_2$ is determined by the invariant mass of the free $\bar q q$ 
system, the argument of $\Psi_g$ is given by the invariant mass of the 
 free $\bar q q g$ system. 
The contribution 
proportional to the  overlap of valence wave functions
 is proportional to the vertex form factor, $T^\Gamma$. 
 The UV  divergence of $T^\Gamma$  
 is canceled 
 by a half of the sum of self-energy corrections  to  two quark lines 
 connected to this vertex. The total contribution 
 to $F_\pi$ from these 
 two self-energy diagrams  can also be cast in the form of Eq.~(\ref{vertex}) with 
 $I_{gg} = 0$ and is  given by 

\begin{eqnarray}
& &  F_\pi(Q^2) \to F^\Sigma_\pi(Q^2) = 
\int {{dx d\k_\perp}\over {16\pi^3}} 
\Psi(x,\k_\perp)T^\Sigma\Psi(x,\k_\perp - x\q_\perp) \nonumber \\
& & +\int {{dx d\k_\perp}\over {16\pi^3}} 
 {{dy d\l_\perp}\over {16\pi^3}} 
 \left[ \Psi_g(x,\k_\perp;y,\l_\perp)T_{g1}^\Sigma \Psi(x,\k_\perp-x\q_\perp)
 +  \Psi(x,\k_\perp)T_{g2}^\Sigma\Psi_g(x,\k_\perp-x\q_\perp;
\l_\perp-y\q_\perp)
 \right], \nonumber \\
\label{self}
\end{eqnarray}
with 
\begin{equation}
 T^\Sigma = -\, 8\pi C_F \alpha_s \int {{dy d\l_\perp}\over {16\pi^3}}
{1\over {y(1-y)}}
 \left[{y \over {D(y,\l_\perp)}} + {y \over {D(y,\l_\perp-y\q_\perp)}}\right], 
\label{sigma}
\end{equation}
and
\begin{equation}
 T_{g1}^\Sigma =   {{\sqrt{8\pi C_F \alpha_s}}\over {y(1-y)}}
  {y\over {D(y,\l_\perp)}},\;\;
 T_{g2}^\Sigma =   {{\sqrt{8\pi C_F \alpha_s}}\over {y(1-y)}}
 {y\over {D(y,\l_\perp-y\q_\perp)}}.
\end{equation}
The remaining UV divergence coming from these two self-energy diagrams  cancel the UV 
 divergences resulting from diagrams which dress the  
axial current  vertices. The IR divergence of $T^\Gamma$ corresponding to 
either $|\l_\perp| \to 0$ or $|\l_\perp - y\q_\perp| \to 0$ is removed by 
 two  contributions in Eq.~(\ref{vert}) proportional to the $\bar q q g$ 
wave functions. When both denominators in $T^\Gamma$ vanish, $i.e.,$   
 $|\l_\perp| \to 0$ and  
$|\l_\perp - y\q_\perp| \to 0$,  all particles in the loop dressing the 
 electromagnetic vertex go on-shell and this singularity 
 is removed by the contribution proportional 
to the $\bar q q g$ wave functions in both initial and final state. 
Similarly, the IR divergent contribution to $T^\Sigma$  in  
Eq.~(\ref{sigma})   
corresponding to either  $|\l_\perp| \to 0$ or $|\l_\perp - y\q_\perp| \to 0$ 
 is canceled by mixing with the $\bar q q g$ wave 
functions.

Consider now the third and fourth diagram shown in Fig.~1. 
 After integrating over the ``minus'' components 
of the loop momenta 
they can also be written in 
the  
form given by Eq.~(\ref{vertex}). 
 In particular, in the limit $|\q_\perp| >> \beta$ the  
 dominant contribution  
 comes from the first term in Eq.~(\ref{vertex}),  
 $ i.e.,$  the overlap of the valence wave 
functions and it  is given by

\begin{equation}
F_\pi(Q^2) \to F_{\pi,{\rm exch}}(Q^2) = 
 \int {{dx d\k_\perp} \over {16\pi^3}}
{{dy d\l_\perp}\over {16\pi^3}} \Psi(y,\l_\perp)
T_{\rm exch}(y,\l_\perp;x,\k_\perp)
\Psi(x,\k_\perp-x\q_\perp), \label{preass}
\end{equation}
where 

\begin{equation}
 T_{\rm exch} = {{16\pi C_F\alpha_s}\over {y-x}} \left[
{{ x(1-x)\l_\perp^2 + y(1-y)\k_\perp^2 + (yx+(1-y)(1-x))\l_\perp\cdot\k_\perp
}
\over { \left({{\l_\perp^2}\over y} + {{(\l_\perp-\k_\perp)^2}\over {x-y}} 
       - {{\k_\perp^2} \over x}\right)
    \left( {{\l_\perp^2}\over {y(1-y)}} - {{\k_\perp^2}\over {x(1-x)}}
 \right)}}\right]. \label{texch}
\end{equation}
To leading order in $1/\q_\perp^2$  this would naively give 

\begin{equation}
{{ 16\pi C_F\alpha_s }\over {\q_\perp^2}}
\int {{dx d\k_\perp}\over {16\pi^3}}
{{dy d\l_\perp}\over {16\pi^3}}
{{\Psi(y,\l_\perp)\Psi(x,\k_\perp)} \over {xy}},  
 \label{assym}
\end{equation}
and therefore the expression for $F_{\pi,{\rm exch}}$ becomes 
 identical to the leading order formula, Eq.~(\ref{fpiass})  with the 
distribution amplitudes 
  given by the asymptotic formula
\begin{equation}
\phi(x) = \int {{d\k_\perp}\over {16\pi^3}}
\Psi(x,\k_\perp) = {{\sqrt{6}}\over 2} f_\pi x(1-x). 
\label{phiass}
\end{equation}
The second equality  above  
 follows from the normalization condition given in 
 Eq.~(\ref{fpi}). Of course Eq.~(\ref{assym}) does not exactly follow from 
 Eq.~(\ref{preass}) since the latter, and therefore also $F_{\pi, {\rm exch}}$,
  is not well defined; $T_{\rm exch}$ has IR divergences  
 coming from  two energy denominators in the square brackets 
in Eq.~(\ref{texch}).  
 These singularities correspond to open $\bar q q$ and $\bar q q g$ 
 intermediate states. These singularities   
 are canceled in Eq.~(\ref{vertex}) by   
  contributions from  $\Psi \otimes T_{1,2g} \otimes  \Psi_g$ and 
 $\Psi_{g} \otimes I_{gg} \otimes \Psi_{g}$, respectively. 
 From $T_{\rm exch}$ we may however define a {\it scheme}-dependent 
 hard contribution 
 kernel,  $T^{\rm hard}_{\rm exch}(\mu)$ and the corresponding hard form 
 factor from one gluon exchange,
\begin{equation}
F_\pi(Q^2) \to F_{\pi,{\rm exch}}^{\rm hard}(Q^2,\mu) =  
  \int {{dx d\k_\perp} \over {16\pi^3}}
{{dy d\l_\perp}\over {16\pi^3}} \Psi(y,\l_\perp)
T_{\rm exch}^{\rm hard}(y,\l_\perp;x,\k_\perp;\mu)
\Psi(x,\k_\perp-x\q_\perp), \label{fexchhard}
\end{equation}
 by cutting-off light cone energy denominators such that 
Eq.~(\ref{preass}) $F_{\pi,{\rm exch}}^{\rm hard}$ 
  becomes IR finite. For example,  we may define  
\begin{equation}
  T^{\rm hard}_{\rm exch}(\mu) \equiv 
\left[\prod_{i=1}^{2}\Theta(D_i)\right]T_{\rm exch}, \;\;
 T^{IR}_{\rm exch}(\mu) \equiv T^{exch} - T^{\rm hard}_{\rm exch} = 
  \left[1 - \prod_{i=1}^2\Theta(D_i)\right]T_{\rm exch},
\end{equation}
where $D_i$ is either one of the two denominators in Eq.~(\ref{texch})
 and  $\Theta(D) \sim 1$,  in the momentum region 
 in which $|D| \agt \mu^2$ and $\Theta(D) \to 0$ for $|D| << \mu^2$.   
The hard contribution from one gluon exchange, $T^{\rm hard}_{\rm exch}$, defined in 
this way is IR finite but cut-off, $\mu$ 
 dependent. 
 The $\mu$-dependent IR singular piece of $T_{\rm exch}$, 
$T^{IR}_{\rm exch}(\mu)$ when combined with 
 the contributions 
 proportional to the $\bar q q g$ wave function produces IR finite  
 but low momentum-dominated contribution. Of course,  when this term is  
combined with  the contribution from $T^{\rm hard}_{\rm exch}(\mu)$,  
the $\mu$-dependence 
disappears.  
 If we choose the factorization scale $\mu$ to be  
 comparable to that setting the   width  of the soft wave function, {\it i.e.} $\mu \sim \beta$
  (or $\mu^2 \sim s_0$ for the local duality wave functions) 
 then  contribution from nonvalence sectors becomes reduced by the 
 residual 
 contribution from  $T^{IR}_{\rm exch} = T_{\rm exch}-T^{\rm hard}_{\rm exch}$, 
 {\it i.e.} the IR dominated 
 piece of the valence contribution. 
 Then for $|\q_\perp| >> \mu$ bulk of the form factor will come from the 
hard gluon exchange  given by $T^{\rm hard}_{\rm exch}$. With $\Theta$ given by  

\begin{equation}
 \Theta(D) = {{ |D| }\over {|D| + \mu^2}}, \label{theta}
\end{equation}
the effect of the nonvalence contributions is to effectively add a 
mass term of order $\mu$ to the two energy denominators, $D_i$ in 
Eq.~(\ref{texch}) which leads to a similar effect as the gluon mass regulator 
 used in Ref.~\cite{JPS}.

\section{Sudakov Suppression}
 
Before proceeding to the analysis of the numerical results, let us 
discuss how the Sudakov suppression from the gluon radiation of the 
struck quark results in our formalisms. 
The exchange of the gluon across the electromagnetic vertex, Eq.~(\ref{qcd2})
gives for the quark spectral function, $\rho^{\rm quark}(s,s')$ 
\begin{eqnarray}
\rho_{\rm vert}(s,s') & = & 2(8\pi C_F\alpha_s) (2\sqrt{6})^2
\int {{dx d\k_\perp}\over {16\pi^3}}
     {{dy d\l_\perp}\over {16\pi^3}}
{
 { y[\l_\perp - \q_\perp]\cdot[\l_\perp + (1-y)\q_\perp] }
 \over {\l_\perp^2[\l_\perp-y\q_\perp]^2} }
 \nonumber \\
& & \times \left[ 
\delta\left(s -  {{\k_\perp^2}\over {x(1-x)}}\right) 
- \delta\left(s - {{\k_\perp^2}\over {x(1-x)}}
                  - {{\l_\perp^2}\over {y(1-y)(1-x)}}\right)
           \right] \nonumber \\
& & \times
           \left[ \delta\left(s' - 
   {{(\k_\perp-x\q_\perp)^2}\over {x(1-x)}} \right)
 - \delta\left(s' - {{(\k_\perp-x\q_\perp)^2}\over {x(1-x)}}
                  - {{(\l_\perp-y\q_\perp)^2}\over {y(1-y)(1-x)}} \right)
            \right].
\label{e1}
\end{eqnarray}
We now write the first term in brackets as

\begin{eqnarray}
& & \left[\delta\left(s -\cdots\right) - \delta\left(s -\cdots\right)\right]
 = 
\delta\left(s -  {{\k_\perp^2}\over {x(1-x)}}\right) 
{ { { {\l_\perp^2}\over {y(1-y)(1-x)}}} \over { {{\l_\perp^2}\over {y(1-y)(1-x)}}
 + {{\k_\perp^2}\over {x(1-x)}}}}
\nonumber \\
& & + 
\left[
\delta\left(s -  {{\k_\perp^2}\over {x(1-x)}}\right) 
{ { {{\k_\perp^2}\over {x(1-x)}}}
 \over { {{\l_\perp^2}\over {y(1-y)(1-x)}} + {{\k_\perp^2}\over {x(1-x)}}}}
- \delta\left(s - {{\k_\perp^2}\over {x(1-x)}}
                  - {{\l_\perp^2}\over {y(1-y)(1-x)}}\right)
\right], \label{e2}
\end{eqnarray}
and similarly for the second term with $s\to s'$ and $\k_\perp
 \to \k_\perp - x\q_\perp$, $\l_\perp \to \l_\perp - y\q_\perp$. 
 In the first term on the right-hand side in Eq.~(\ref{e2}) 
 the $q\bar q g$ invariant mass, 
$\l_\perp^2/(y(1-y)(1-x))$ in now regularized in the IR by the incoming 
$q\bar q$ virtuality, $\k_\perp^2/x(1-x)$. Thus when integrated in 
 Eq.~(\ref{e1})
 over gluon transverse momentum the first term 
 in Eq.~(\ref{e2}) gives an IR finite contribution. 
 The term in the bracket in Eq.~(\ref{e2}) is also IR finite. 
 However,  the two delta  functions give separately IR singular
 contributions [from integration over $1/\l_\perp^2$].  
 After integrating over $s$ [with $e^{-s/2\beta^2}$
for the model based on the Borel  transformation
 or $\theta(s_0 - s)$ 
  for that  suggested by  the  local duality], 
 these become  proportional to the valence $q\bar q$ and $q \bar q g$ 
 wave functions, respectively. 
 Combining the IR finite contribution from the pure $\bar q q$ sector 
(first term in the right-hand side  of Eq.~(\ref{e2}) and a similar one 
 for $s\to s'$) 
 gives 
\begin{eqnarray}
& & \rho^{\rm hard}_{\rm vert}(s,s')  =  2(8\pi C_F\alpha_s) (2\sqrt{6})^2
\int {{dx d\k_\perp}\over {16\pi^3}}
     {{dy d\l_\perp}\over {16\pi^3}}
\delta\left(s -  {{\k_\perp^2}\over {x(1-x)}}\right) 
\delta\left(s' - 
   {{(\k_\perp-x\q_\perp)^2}\over {x(1-x)}} \right) \nonumber \\
& & \times {
 { y[\l_\perp - \q_\perp]\cdot[\l_\perp + (1-y)\q_\perp] }
 \over {\left[ \l_\perp^2 +  y(1-y){{\k_\perp^2}\over {x}} \right]
\left[ (\l_\perp-y\q_\perp)^2
 + y(1-y){{(\k_\perp - x\q_\perp)^2}\over {x}} \right]}}.
\end{eqnarray}
Treating similarly two self-energy contributions for the struck quark, 
  and adding them  to  the vertex corrections gives  
 the total   IR finite electromagnetic current $q\bar q$ matrix element 

\begin{equation}
\rho^{\rm hard}_{{\rm em}}(s,s') = S(x,\k_\perp,\q_\perp)
(2\sqrt{6})\delta\left(s - {{\k_\perp^2}\over {x(1-x)}}\right)
(2\sqrt{6})\delta\left(s' - {{(\k_\perp-x\q_\perp)^2}\over {x(1-x)}}\right), 
\end{equation}
where the form factor $S$ is given by 
\begin{equation}
S(x,\k_\perp,\q_\perp) = 
1  
- 4\pi C_F\alpha_s 
\int {{dy d\l_\perp}\over {16\pi^3}} y 
{{ \left[ \left(2(1-y) + y^2\right) \q_\perp^2 
+ y(1-y) \left( {{\k_\perp^2}\over x} + 
 {{(\k_\perp- x \q_\perp)^2}\over x} \right) \right]}
\over {
\left[ \l_\perp^2 +  y(1-y){{\k_\perp^2}\over {x}}\right]
\left[ (\l_\perp-y\q_\perp)^2
 + y(1-y){{(\k_\perp - x\q_\perp)^2}\over {x}} \right]}}. \label{sud1}
\end{equation}
The contribution from this spectral function to $F_\pi$ is thus given by 
\begin{equation}
F_\pi(Q^2) \to F_{\pi,{\rm em}}^{\rm hard}(Q^2) = 
\int [dxd\k_\perp] \Psi(x,\k_\perp)S(x,\k_\perp,\q_\perp)
\Psi(x,\k_\perp-x\q_\perp). 
\end{equation}
Due to the wave function suppression, both $\k^2_\perp/x(1-x)$ and 
$(\k - x\q_\perp)^2/x(1-x)$ are small as compared to $\q_\perp^2$ 
 in the integral in Eq.~(\ref{sud1}). Thus, the integral for the 
 Sudakov form factor is dominated by  
  $\l_\perp,y\to 0$ and 
$\l_\perp -y\q_\perp,y\to 0$. In these two regions, both transverse  
$\l_\perp$  
and longitudinal  $y$ momentum integration lead to $\log(\q_\perp^2)$.  
To leading order in  $(\k^2_\perp/x(1-x))/\q_\perp^2$ and 
$((\k - x\q_\perp)^2/x(1-x))/\q_\perp^2$, the result is 

\begin{equation}
S(x,\k_\perp,\q_\perp) \approx 1 - 
{{C_F\alpha_s}\over {4\pi}}\log^2
\left( {{x\q_\perp^2}\over {\k_\perp^2}}\right)
-{{C_F\alpha_s}\over {4\pi}}\log^2\left( {{x\q_\perp^2}\over 
{(\k_\perp-x\q_\perp)^2}}\right). \label{sud}
\end{equation}

Alternatively, we may define Sudakov form factor introducing factorization 
scale, $\mu$ to regularize the small virtuality of $q\bar q g$ 
intermediate state, just as  
 we did in defining hard contribution from the one gluon exchange diagrams.  
We thus rewrite the differences of the two $\delta$-functions in Eq.~(\ref{e1}) as

\begin{eqnarray}
& & \left[\delta\left(s - \cdots\right) - \delta\left(s - \cdots\right)\right]
 = 
\delta\left(s -  {{\k_\perp^2}\over {x(1-x)}}\right) 
\Theta\left(  {{\k_\perp^2}\over {x(1-x)}} \right)
\nonumber \\
& & + 
\left[
\delta\left(s -  {{\k_\perp^2}\over {x(1-x)}}\right)\left(1 - \Theta\left(
  {{\k_\perp^2}\over {x(1-x)}} \right) \right)
- \delta\left(s - {{\k_\perp^2}\over {x(1-x)}}
                  - {{\l_\perp^2}\over {y(1-y)(1-x)}}\right)
\right], \label{e22}
\end{eqnarray}
with  $\Theta$ defined in Eq.~(\ref{theta}), 
and similarly for the $s'$-dependent part. 
 The hard contribution which defines  perturbative Sudakov form factor 
comes, as before, from the first term on the  right-hand side.  
Collecting hard contribution from vertex and 
   self energies results in a contribution to $F_\pi$ given by 

\begin{equation}
F_\pi(Q^2) \to F_{\pi,{\rm em}}^{\rm hard}(Q^2,\mu)  = 
 \int {{dx d\k_\perp}\over {16\pi^3}} \Psi(x,\k_\perp)S(x,\k_\perp,\q_\perp;\mu)\Psi(x,\k_\perp-x\q_\perp), \label{vertexhard}
\end{equation}
where $S(x,\k_\perp,\q_\perp;\mu)$ is now given by Eq.~(\ref{sud1}) with 
$\k_\perp^2/x$ and $(\k_\perp -x\q_\perp)^2/x$ in the denominator replaced by 
 $(1-x)\mu^2$. For $\mu^2 \sim \beta^2 << \q_\perp^2$ the $\mu$-dependent 
 Sudakov form factor is thus given by 

\begin{equation}
S(x,\k_\perp,\q_\perp;\mu) 
\approx 1 - 
{{C_F\alpha_s}\over {2\pi}}\log^2
\left( {{\q_\perp^2}\over {(1-x)\mu^2}}\right). \label{sudmu}
\end{equation}
The full contribution from the diagram on Fig.~1b including self energy 
corrections becomes

 \begin{eqnarray}
 & & F_\pi(Q^2) \to F_{\pi,{\rm em}}(Q^2,\mu)  = 
 \int {{dx d\k_\perp}\over {16\pi^3}} \Psi(x,\k_\perp)S(x,\k_\perp,\q_\perp;
\mu)]
\Psi(x,\k_\perp-x\q_\perp)  + \int {{dx d\k_\perp}\over {16\pi^3}}
{{dy d\l_\perp}\over {16\pi^3}}
 \nonumber \\
& & \times
\left[\hat\Psi_g(x,\k_\perp;y,\l_\perp)T_{g1}^{\rm hard}\Psi(x,\k_\perp-x\q_\perp)
 + \Psi(x,\k_\perp)T_{g2}^{\rm hard}
\hat\Psi_g(x,\k_\perp-x\q_\perp;y,\l_\perp-y\q_\perp) \right.\nonumber \\
& &  \phantom{  F_\pi^\Gamma(Q^2) = \int {{dx d\k_\perp}\over {16\pi^3}}{{dy
d\l_\perp}\over {16\pi^3}}
} \left.
  +\hat\Psi_g(x,\k_\perp;y,\l_\perp)
I_{gg}^\Gamma\hat\Psi_g(x,\k_\perp-x\q_\perp;y,\l_\perp-y\q_\perp)\right].
\label{verth}
\end{eqnarray}
Introduction of the factorization scale modifies the $T_{gi}$ amplitudes 
 according to 

\begin{equation}
T_{gi} \to T_{gi}^{\rm hard}  = {{\sqrt{8\pi C_F \alpha_s}} \over {y(1-y)}}
\left[ {y\over 2} - N(y,\l_\perp){{\Theta(D_i)}\over {D_i}}\right],\;\; (i=1,2)
\end{equation}
and the modification of the nonvalence wave functions $\Psi_g \to \hat\Psi_g$ 
comes from the subtraction of the IR singular part 
 of the valence sector,

\begin{equation}
\hat \Psi_g = \Psi_g - {{\sqrt{8\pi\alpha_sC_F}[1 - \Theta(D_i)]}\over {D_i}}
\Psi. \label{peff}
\end{equation}
 Even though each individual term in 
Eq.~(\ref{verth}) is $\mu$-dependent the entire sum is $\mu$-independent.
For $\mu^2 \sim \beta^2 \sim \langle \k_\perp^2/x(1-x)\rangle << \q_\perp^2$ 
the second term in Eq.~(\ref{peff}) strongly reduces 
the nonvalence amplitudes, $\hat \Psi_g$ and the vertex correction becomes 
approximately given by the valence hard contribution alone, {\it i.e.} 
$F_{\pi,{\rm em}}^{\rm hard}(Q^2,\mu\sim\beta)$. Also in this case the two 
expressions for the Sudakov form factor, Eqs.~(\ref{sud}) and 
(\ref{sudmu}) give almost the same result.

In the ladder approximation Sudakov logarithms exponentiate and 
$S$ becomes,

\begin{equation}
S \sim \exp\left( 
  - {{\alpha_s}\over {2\pi}}C_F\log^2  {{Q^2}\over {(1-x)\mu^2}} 
  \right).\label{sudall}
\end{equation}
To get rid of the $\mu$ dependence in this case one should 
 sum to all orders 
 in $\alpha_s^n$ 
contributions from $q \bar  q g^n$ wave functions whose IR singularities 
 cancel by mixing with nearby Fock sectors. However, as discussed above 
 for the $n=1$ case, if 
 $\mu \sim \beta$ this 
 residual contribution is expected to be small.

\section{Numerical Results}

The numerical results are summarized in Figs.~2-4. 
 They are all given for fixed  $\beta = 0.4 \mbox{ GeV}$.

 In Figs.~2a,b,
 corresponding to $\alpha_s=0.25$ and $0.4$ respectively, 
 the triangles represent  the contribution $F_\pi^0$ of  the
lowest-order diagram Fig.~1a, i.e., the 
purely soft contribution determined by the Gaussian wave function of 
Eq.~(\ref{wfpi}). At high $Q^2$, the function  $F_\pi^0(Q^2)$ 
falls off like $1/Q^4$. 
 For comparison, in Fig.~2a we also show 
 the results of calculation of the 
soft contribution  within  the QCD sum rule method~\cite{IOSMNU}
 (dashed line). 
The set of 
 circles represents the sum $F_\pi^0 + F^\Gamma_\pi + F^\Sigma_\pi$
of the lowest-order  contribution and the $O(\alpha_s)$ 
correction to the  electromagnetic vertex 
 $F^\Gamma_\pi$ (\ref{vert}) accompanied by the 
quark self-energy correction   $F^\Sigma_\pi$ (\ref{self}). 
As shown in Sec.III, the exponentiation of the leading double-logarithmic
terms suppresses vertex correction and it can be approximated by  
 $F^{\rm hard}_{\pi,em}$ given in Eq.~(\ref{vertexhard}). Using $S$ from  
 Eq.~(\ref{sudall}) with $\mu^2 = 2\beta^2$, 
we obtain the curve   depicted by the squares.
Finally, the solid line $F^{\rm tot}_{\pi}$ 
combines $F^{\rm hard}_{\pi,em}$ with the remaining 
 contribution from the one gluon exchange diagrams Figs.~1c,d,   
 the remaining half of the struck quark self-energy diagrams
 and the spectator's self-energy diagram.

Fig.~3 gives the comparison of the form factor calculation for two model 
wave functions discussed in Sec.II. The upper and lower dashed lines represent 
 the soft  contribution from Eq.~(\ref{soft}) for the wave functions given
 by Eq.~(\ref{wfpi}) and Eq.~(\ref{pld}), respectively. 
The upper and lower solid lines are the
 respective results for $F^{\rm tot}_{\pi}$ (i.e.,  full 
 calculation to $O(\alpha_s)$ for gluon exchange diagrams and 
Sudakov-exponentiated 
vertex correction; 
 $\alpha_s=0.25$ was used). 
The overall model dependence is seen to be rather small. 
 As discussed in Sec. II, the  one gluon  exchange 
 diagrams are responsible for the  enhancement of the 
 hard contribution at high $Q^2$.

 Inspecting the results shown in Fig.~2a (2b)  
  we may conclude that for $Q^2 \sim \mbox{few GeV}^2$ the dominant effect 
 comes from 
  the soft region.  The next-order contribution 
 is $\sim 7\%  $ ($\sim 10\%  $)  at $Q^2 \sim 1\mbox{ GeV}^2$ 
 and increases to about 
 $20\%  $  ($30\%  $) at $Q^2 \sim 4\, \mbox{ GeV}^2$. 
A large fraction of the enhancement from one gluon exchange diagrams 
is however IR dominated. This part of gluon exchange should therefore 
 be considered together with the 
soft wave function contribution and separated from the 
hard scattering amplitude represented by hard gluon exchange. 
This effect is illustrated in Fig.~4. The  
contributions from hard gluon exchange given by $T_{\rm exch}^{\rm hard}$ 
 is plotted together with the remaining contributions from the one gluon 
exchange diagrams. The dashed line is the result of the asymptotic formula (cf. 
Eq.~(\ref{fpiass})) with the asymptotic distribution amplitude given by 
Eq.~(\ref{phiass}). The solid line which approaches the asymptotic result 
from below is the contribution from $T^{\rm hard}_{\rm exch}$ with $\mu^2 = 2\beta^2$. 
The other solid line is the leftover from the one gluon exchange diagrams. 
  The two sets of  points 
 shown by circles and triangles below the asymptotic curve represent 
 the hard gluon exchange contribution for $\mu^2 = 10\beta^2$ and 
$\mu^2 = 0.2\beta^2$, respectively. The upper circles and triangles give 
 then the remaining one gluon exchange contributions for these 
two values of the factorization scale,  respectively. 
The net contribution from each of the three sets of points (solid lines, 
 circles and triangles)  is 
$\mu$-independent.

As advocated previously  
$\mu \sim \beta$ leads to the fastest saturation of the asymptotic form factor 
by the pure hard gluon exchange {\it i.e.} as $Q^2$ increases 
 the choice $\mu \sim \beta$ is optimal for reducing the contribution from 
 nonvalence sectors. 
 At $Q^2 \sim 20\, \mbox{GeV}^2$ the hard contribution starts to dominate over  
 the IR  sensitive part of the one gluon exchange. 
   This is consistent with the results  found in Ref.~\cite{JPS}  
 where 
  sensitivity of the form factor to 
 the cut-off imposed on the light cone energy denominators
 was studied. 
The discussion on the normalization of the space-like
pion form factor data can also be found in the recent analysis 
with the optimal renormalization scale and scheme dependence \cite{bjpr}.

\section{Conclusions}

Our main focus in this work was  to consistently 
generate  the gluon  radiative corrections  within 
an approach motivated by the  light cone quantization
formalism and QCD sum rules.  
Just as in the latter approach, our starting objects are the 
 Borel-transformed 
Green's functions.   To make a link with the 
light-cone quantization, we    demonstrated 
that the action of the Borel transform is analogous to using 
 the Gaussian valence soft wave function 
which was frequently used in the past \cite{bhl}. 
Such a pion wave function  can be made to satisfy  
the standard light cone normalization conditions for  the pion decay 
constant, form factor, etc.
The main advantage of the  Green's functions  method 
is that applying the Borel transformation 
to the two-loop Green's functions, we generate 
both the radiative corrections to the current matrix element 
and  the non-valence ${\bar q}qg$ components of the 
pion wave function which are absolutely  necessary to 
secure  the gauge invariant and infrared-finite 
results for the total $O(\alpha_s)$ corrections to the pion form factor.
In addition, all of these wave functions are readily applicable to the light 
cone time ordered perturbative expansion of pion form factor beyond 
the leading twist level in QCD.  
Further works involving the four-point  Green's functions and 
applications to the
virtual  Compton scattering are in progress.

\vspace{0.5cm}

\centerline{\large \bf Acknowledgment}

\vspace{0.5cm}

We thank  I. Musatov  for useful  discussions. 

This work was supported by the 
US Department of Energy  under contracts DE-AC05-84ER40150, 
DE-FG05-88ER40461, DE-FG02-96ER40947 
 and also by
Polish-U.S. II Joint Maria Sklodowska-Curie
Fund, project number PAA/NSF-94-158.

\begin{figure}[t]
\mbox{
\epsfxsize=5in
\epsfysize=5in
\hspace{1cm} \epsffile{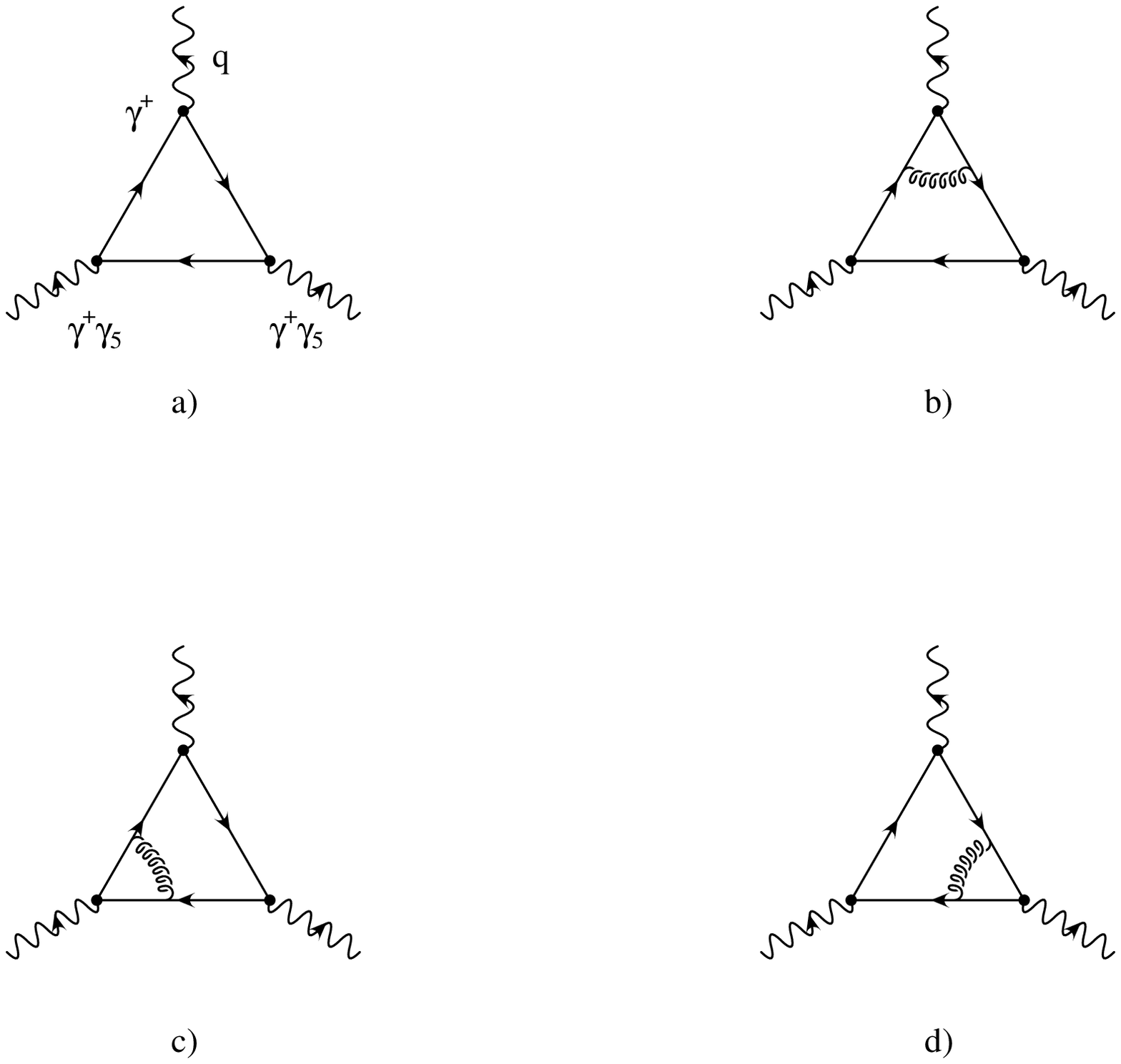} 
}
{\caption{\label{fig:amb1}
 Perturbative expansion of the three point function used to
 calculate the form factor.}}
\end{figure}

\vspace{-5cm}
\begin{figure}[h]
\mbox{
\epsfxsize=5in
\epsfysize=5in
\hspace{-1.5cm} \epsffile{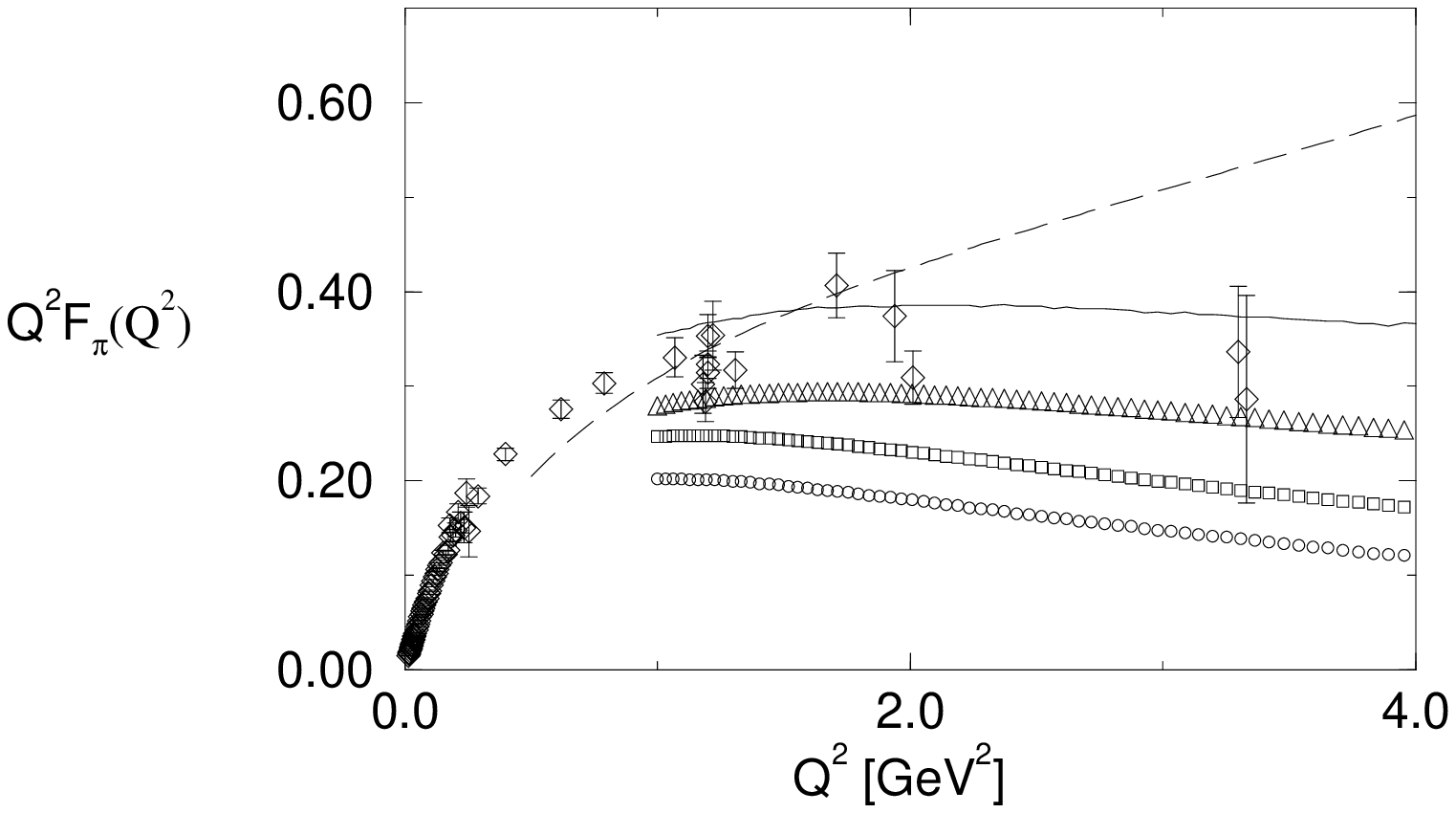} 
\epsfxsize=5in
\epsfysize=5in
\hspace{-4cm}
\epsffile{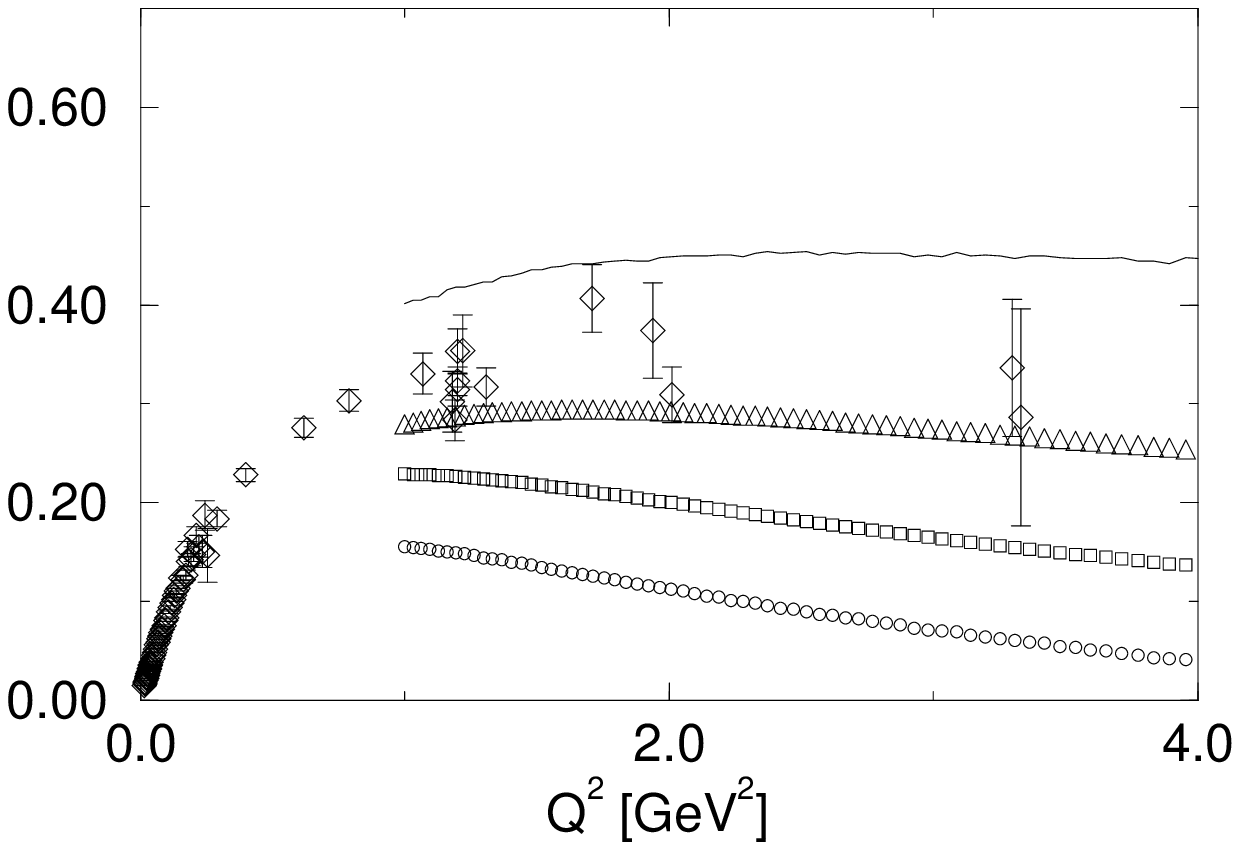} 
}
{\caption{\label{fig:amb2}
 QCD corrections to $Q^2F_\pi(Q^2)$.}}
\end{figure}

\vspace{-4cm}
\begin{figure}[h]
\mbox{
\epsfxsize=5in
\hspace{1cm} \epsffile{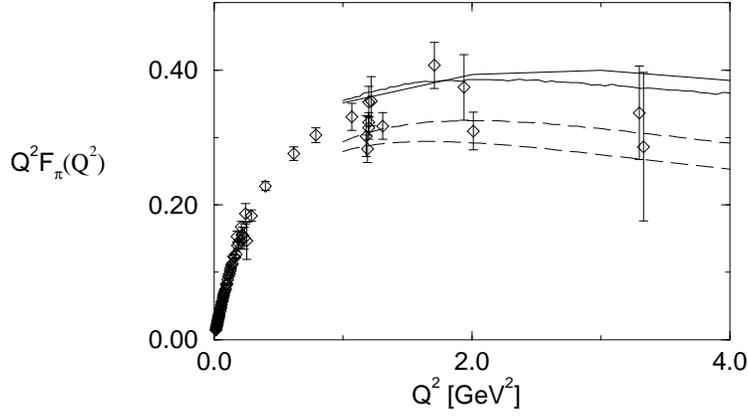} 
}
{\caption{\label{fig:amb3}
Wave function dependence of QCD  corrections to
    $Q^2F_\pi(Q^2)$.}}
\end{figure}

\vspace{-5cm}
\begin{figure}[h]
\mbox{
\epsfxsize=5in
\epsfysize=5in
\hspace{1cm} \epsffile{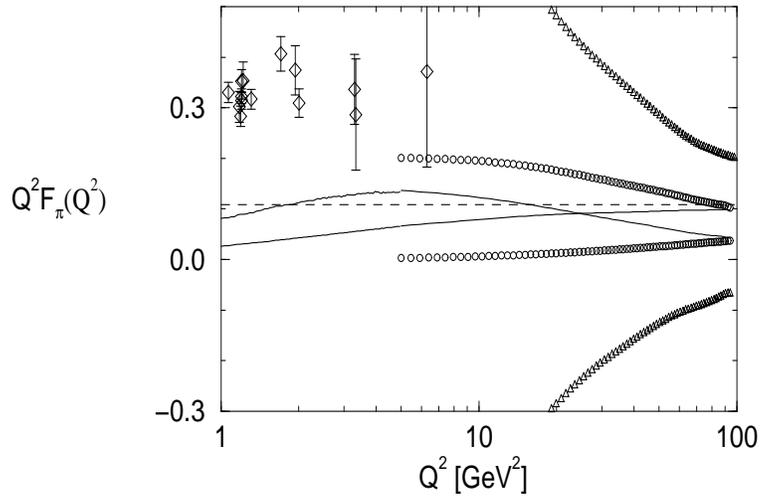} 
}
{\caption{\label{fig:amb4}
Gluon exchange vs. hard gluon exchange  
 contribution to $Q^2 F_\pi(Q^2)$.}}
\end{figure}

\end{document}